\documentstyle[aps,prb]{revtex}

\begin{document}
\draft
\preprint{}
\title{Electronic structure of CdTe nanocrystals: A tight-binding study}
\author{Jes\'us P\'erez-Conde}
\address{Departamento de F{\'\i}sica, Universidad P\'ublica de Navarra,
E-31006 Pamplona, Spain}
\author{A. K. Bhattacharjee}
\address{Laboratoire de Physique des Solides, URA au CNRS,
Universit\'e Paris-Sud, 91405 0rsay, France}
\date{\today}
\maketitle
\begin{abstract}
We present a symmetry-based calculation of the electronic structure
of a compound semiconductor quantum dot (QD) in the $sp^3s^*$
tight-binding model
including the spin-orbit interaction. The Hamiltonian matrix is diagonalized
exactly for CdTe QD sizes up to 60 {\AA}. The surface dangling bonds
are passivated by hydrogen through a careful analysis of the density of states
and wave functions. The calculated size dependence of the energy gap
shows a reasonable agreement with the available experimental data. Our symmetry
analysis indicates that, in contrast with a reported prediction of the
three-band effective-mass model, the fundamental interband transition
remains dipole-allowed in CdTe nanocrystals.
\end{abstract}


\narrowtext

\section{Introduction}

Semiconductor nanocrystallites, also called quantum dots (QD's), have been
studied in recent years.\cite{wang}
In particular, high-quality crystallites
of II-VI compounds have been fabricated
and their optical properties investigated in detail.\cite{bawendi,masumoto}
The confinement-induced
blueshift of the fundamental gap and the discretization of the energy
spectrum have been observed.
Early theoretical studies \cite{efros,kayanuma} were based on the effective
mass approximation (EMA),
which was progressively refined by taking into
account the degeneracy of the valence band.\cite{xia,grigoryan,sercel,lefevre}
Generally speaking, the EMA overestimates the confinement energy in small
QD's. Lippens and Lannoo\cite{lippens} reported a tight-binding (TB)
calculation
of the QD energy gap based on the empirical $sp^3s^*$ model, previously
developed
by Vogl {\it et al.}\cite{vogl} for bulk semiconductors. They used the
recursion method and
obtained a better agreement with experimental data than the EMA
calculations. Nair {\it et al.}\cite{nair}
presented a different TB approach based on an effective-bond-orbital
model.\cite{chang}
The pseudopotential method has also been succesfully applied to
some QD's.\cite{rama,zunger}

Ren and Dow\cite{ren} developed an algorithm for exact diagonalization
of the TB Hamiltonian for a hydrogenated Si cluster in the $sp^3s^*$ model by
using the tetrahedral ($T_d$) symmetry.
More recently, Albe {\it et al.}\cite{albe} reported TB calculations
for ZnS and CdSe QD's, also based on an exact diagonalization of the cluster
Hamiltonian, but {\it including} the spin-orbit interaction. However,
the ``brute force'' method, suitable for studying the QD shape effects they
focused on, is rather limited to small sizes and does not allow any symmetry
analysis of the QD eigenstates.

In this paper we report a symmetry-based TB study of the electronic structure
of a zincblende semiconductor crystallite of roughly spherical shape
in the $sp^3s^*$ model including the spin-orbit interaction. Thus, our approach
generalizes that of Ref. \onlinecite{ren} to binary compounds and a finite
spin-orbit interaction.
Here we apply it to CdTe, because substantial experimental
data\cite{masumoto,rajh,mastai} are available
and a thorough three-band EMA study\cite{lefevre} has been reported,
allowing a detailed comparison with our results.

The paper is organized as follows. We first present an outline of the
model in Section II, including the method of
passivation of the surface states through hydrogen bonding. We
present and discuss some results for CdTe QD's in Section III.
The size dependence of the energy
and the symmetry classification of a few conduction and valence band edge
states are shown. The variation of the energy gap is compared with the
available experimental
data and the results of other published calculations. Finally, an
analysis of the orbital symmetry of the highest valence band
and
the lowest conduction band states is carried out in order to check
the electric dipole selection rule for the fundamental transition.
The paper closes with some concluding remarks.

\section{The model}

The spin-independent part of the TB Hamiltonian is given
by,\cite{vogl,kobayashi}
\begin{equation}
 H_{0}=
\sum_{b,{\bf R}_{b},i,\sigma}
\vert{{\bf R}_{b},i,\sigma}\rangle E_{i,b}\langle{{\bf R}_{b},i,\sigma}\vert+
\sum_{<{\bf R}_{a},{\bf R}_{c}>,i,j,\sigma}
 \vert{{\bf R}_{a},i,\sigma}\rangle V_{i,j}\langle{{\bf R}_{c},j,\sigma}\vert+
 \mbox{H.c.},
\end{equation}
where $b=a$ (anion), $c$ (cation).
${\bf R}_{b}$ are the atomic position vectors. $i$ and $j$ denote the
orthonormal atomic orbitals $s$, $p_x$, $p_y$, $p_z$ and $s^{*}$, which depend
on $b$. $\sigma$ is the $z$
component of spin ($\uparrow$ and $\downarrow$). Note that $s^*$ represents an
an excited $s$ state introduced phenomenologically in order to obtain a correct
description of the conduction band in bulk semiconductors.
The interatomic matrix elements $V_{i,j}$ are restricted to the nearest
neighbors as
indicated by $<{\bf R}_{a},{\bf R}_{c}>$ in the summation index. 13
independent parameters then characterize $H_0$ for a compound
semiconductor, which are chosen to fit the known band structure. For
example, the parameters for CdTe are given by\cite{kobayashi} $E_{s,a}=-8.891$,
$E_{p,a}=0.915$, $E_{s,c}=-0.589$, $E_{p,c}=4.315$, $V_{s,s}=-4.779$,
$V_{x,x}=2.355$, $V_{x,y}=4.124$, $V_{s,p}=1.739$, $V_{p,s}=-4.767$,
$E_{s^*,a}=7.0$,
$E_{s^*,c}=7.5$, $V_{s^*,p}=1.949$, $V_{p,s^*}=-2.649$ eV.
The spin-orbit coupling
part of the Hamiltonian mixes the spin-up and spin-down $p$ orbitals on the
same atom:
\begin{equation}
H_{\text{SO}}=\sum_{b,{\bf R}_{b},\sigma,\sigma^{\prime},i,j}
\vert{{\bf R}_{b},i,\sigma}\rangle 2\lambda_{b}{\bf L}_{b}\cdot{\bf S}_{b}
\langle{{\bf R}_{b},j,\sigma^{\prime}}\vert.
\end{equation}
It introduces 2 additional parameters. $\lambda_a = 0.367$ and
$\lambda_c = 0.013$ eV for CdTe.\cite{kobayashi}
When $H_{\text{SO}}$ is included a more convenient atomic basis is given by the
spin-orbit coupled orbitals, which are basically
the total angular momentun eigenstates ${\vert{\bf R}_b,j,j_z}\rangle$
($j=3/2,1/2$) that diagonalize the spin-orbit interaction. We, however,
write them in terms of the basis functions 
 $\vert{{\bf R}_b,u^k_m}\rangle$
of the irreducible representations $\Gamma_k$ ($k=6, 7, 8$) of the
tetrahedral
point group $T_d$ with respect to the site ${\bf R}_b$. Note that we have two
sets of $\Gamma _6$ orbitals at a given site arising from the $s$ and $s^*$
states. Using the phase convention and coupling coefficients
of Ref. \onlinecite{koster}, the first rows are
\[
\begin{array}{lc}
 \vert{u^{6}_{-1/2}(s)}\rangle=
 \vert{s}\rangle \vert{\downarrow}\rangle, \\
 \vert{u^{7}_{-1/2}(p)}\rangle=
 -\frac{i}{\sqrt{3}}\vert{p_x}\rangle
 \vert{\uparrow}\rangle
  -\frac{1}{\sqrt{3}}\vert{p_y}\rangle
 \vert{\uparrow}\rangle
  +\frac{i}{\sqrt{3}}\vert{p_z}\rangle
  \vert{\downarrow}\rangle, \\
  \vert{u^{8}_{-3/2}(p)}\rangle=
 -\frac{i}{\sqrt{6}}\vert{p_x}\rangle
 \vert{\downarrow}\rangle
  +\frac{1}{\sqrt{6}}\vert{p_y}\rangle
  \vert{\downarrow}\rangle
  +\frac{i\sqrt{2}}{\sqrt{3}}
 \vert{p_z}\rangle\vert{\uparrow}\rangle.
\end{array}
\]

Let us now construct a cluster of roughly spherical shape
starting from, say, a cation at the origin by successively adding
nearest-neighbor atoms through tetrahedral bonding. The dangling
bonds emanating from the atoms generated in the last step will be passivated by
placing a hydrogen $s$ orbital at each empty nearest-neighbor site. This
``hydrogenated'' crystallite has an overall tetrahedral
symmetry. We can, therefore, reduce the Hamiltonian to a block diagonal form
by rewriting it in a symmetrized basis corresponding to the double-valued
representations $\Gamma_k$ ($k=6, 7, 8$) of $T_d$.

Following the procedure developed
in Ref. \onlinecite{chang}, we first construct the symmetrized site
functions $\phi ^i_n$, corresponding to the single-valued representations
$\Gamma _i (i=1,2,3,4,5)$ of $T_d$.
All atoms at a given distance $R_b$ from the origin constitute a shell.
The sites $\{{\bf R}_b\}$ on a given shell are grouped into (generally
more than one) symmetry subshells, each containing all sites which transform
into one another under the 24 symmetry operations of $T_d$ with respect
to the QD center. Each subshell $\{{\bf R}\}$ is then spanned by the functions
$\phi ^i_n$ which represent symmetrized linear combinations of sites, assuming
the values $\phi ^i_n({\bf R})$ at ${\bf R}$. These values are deduced by using
the projection operators. Note that, except for the one-dimensional
representations $\Gamma _1$ and $\Gamma _2$, an irreducible representation
generally occurs more than once in a given subshell.
The site functions are then coupled with the localized atomic orbitals and the
subshell basis functions of total symmetry $\Gamma _k (k=6,7,8)$ obtained
as linear combinations of $\sum_{\bf R}\phi ^i_n({\bf R})
\vert{{\bf R},u^j_m}\rangle$,
by using
the coupling coefficients for $\Gamma_k$ in $\Gamma _i \times \Gamma_j$.

The hydrogen shells are treated in a similar way. Here we have
only one $s$-orbital per atom so that the local basis is restricted
to the $\Gamma_6$ symmetry: \mbox{$\vert{{\bf R}_{\text{H}},u^{6}_m}\rangle$}.
 Finally, in the symmetrized basis,
the total Hamiltonian becomes block diagonal, each block corresponding to
a given symmetry $\Gamma_{i} (i=6,7,8)$.
This implies a substantial reduction of the size of the matrix to diagonalize.
For example, for a cluster containing 3109
semiconductor atoms plus 852 H,
instead of a matrix
 of size $32794=10\times 3109+2\times 852$ we get two equivalent $\Gamma_6$
 matrices of size $2740$, two equivalent $\Gamma_7$ matrices of size $2739$
 and four equivalent $\Gamma_8$ matrices of size $5459$. Thus, the matrix size
has been reduced by a factor of 6.

While the TB parameters enumerated above are assumed to be the same as those
for the bulk semiconductor, we need to precise the parameters concerning the
hydrogen atoms.
The H energy level is obtained through
the same scaling prescription as that for the cation and
 anion $s$ levels.\cite{kobayashi}
In the case of CdTe,
we obtain
 $E_{s,\mbox{\scriptsize H}}=-5$ eV.
 The hopping matrix elements between the
 anion or cation and H are assumed to follow the Harrison scaling rule:
$V_{b-\text{H}}=(d_{a-c}/d_{b-\text{H}})^2 V_{ac}$,
in terms of the bond lengths. $d_{b-\text{H}}$ are usually\cite{ren} taken from
molecular data. In the absence of such data on Cd-H or Te-H bonds,
we estimate the bond lengths by summing the corresponding covalent radii.
This procedure typically yields the experimental bond lengths to within
a few hundredths of angstrom. We thus obtain
$d_{\text{Cd-H}}=1.71$ and $d_{\text{Te-H}}=1.67$ {\AA}.
The above parametrization scheme is adequate for passivating
the surface dangling bonds in Si and Ge nanocrystals.\cite{ren}
However, in the case of CdTe, as shown in the next section, it is necessary
to readjust (reduce) the bond lengths in order to completely eliminate the
surface
states from the electronic spectrum in the neighborhood of the forbidden
gap.

\section{results and discussion}
Let us first discuss the saturation of dangling bonds, as we
are basically dealing with surface-passivated QD systems which show a
systematic increase of the optical energy gap with decreasing size.
The electronic density of states (DOS) as well as its hydrogen projected part,
as calculated from the exact eigenstates of the TB Hamiltonian,
are partly shown in Fig.~1 for a CdTe QD of
diameter $D=58.67$ {\AA} containing $N=3109$ semiconductor atoms.
Note that $D\equiv a(3N/4\pi)^{1/3}$.
The uppermost spectrum (a) corresponds to the CdTe cluster without H.
Notice the large number of states in the bulk band gap region between 0
and 1.6 eV, which arise from the surface dangling bonds. Next we show
the results (b) for the hydrogenated cluster (852 H atoms) with the standard
bond length parameters:
$d_{\text{Cd-H}}=1.71$ and $d_{\text{Te-H}}=1.67$ {\AA}. Most of
the surface states have disappeared, but not all. In fact, as seen from the
hydrogenic partial DOS superposed on the total DOS, the lowest
unoccupied (conduction) states are still surface states. We have
directly verified this through the spatial distribution of the wave function.
This is in contrast with the cases of Si or Ge where a complete
saturation of dangling bonds is achieved by using the molecular bond lengths
for the couplings with H atoms.
As shown in Fig. 1 (c), in order to eliminate surface states
from the relevant part of the electronic spectrum in our CdTe QD, we
need to increase the couplings with H. By
 choosing bond lenghts slightly shorter than before,
$d_{\text{Cd-H}}=1.58$ and $d_{\text{Te-H}}=1.54$ {\AA},
 the hydrogenic DOS in the conduction band vanishes up to $\sim 2$ eV. We have
checked
that the same set of bond length parameters assures surface passivation in
all the QD's considered.

Now we present the energies and wave functions of CdTe quantum dots of
 several sizes (see Table~\ref{tab1}). We have chosen cation- and
anion-terminated crystallites alternatingly in order to show the typical
oscillating behavior of energy levels
in the clusters of binary compounds.\cite{albe} For instance, two
crystallites
of similar diameters, $12.35$ and $13.15$ {\AA}, and terminated by cation and
  anion
 shells, respectively, present the lowest unoccupied state (LUS) at $2.11$ and
$2.42$ eV.
This is in contrast with the monotonous decrease
 observed in Si or Ge crystallites.
 The valence states also show such oscillations but with a smaller amplitude.
 In general, the LUS energy of a cation-terminated
crystallite is higher than that
 of an anion-terminated crystallite of neighboring size, because of the
dominating contribution from the cation $s$ orbital to the conduction band.
 In Figures \ref{fig2} and \ref{fig3} we plot a few energy levels
below and above the bandgap, respectively, against the QD size.
The symmetry classification of the levels is also indicated.
Note that for all sizes the LUS (conduction band)
is of
$\Gamma_6$ symmetry and the highest occupied state (HOS) of the valence band
belongs to $\Gamma_8$, as in the bulk semiconductor.
For comparison, we include the respective energies calculated in the
EMA (Ref. \onlinecite{lefevre}) which are shown as the solid curves.
As usual, the EMA confinement energies are much larger than the TB ones in
small-size QD's. However, the HOS (Fig. 2) shows a better convergence with
increasing size, which reminds us that the TB model yields a better
description of the valence band in the bulk semiconductor.

 The calculated QD bandgap $E_{g}$,
 the energy difference between the LUS and HOS,
 is plotted as a function of size in
  Fig.\ \ref{fig4}. We also present a compilation of the available experimental
values of the optical bandgap. The results of previous EMA (Ref.
\onlinecite{lefevre}) and TB (Ref. \onlinecite{lippens}) calculations
are also shown. In contrast with these calculations, however, we have not
corrected $E_{g}$ for the electron-hole Coulomb interaction. Thus,
the closeness
of our results to those of Lippens and Lannoo\cite{lippens}
shown as the dashed curve is rather misleading.
Their uncorrected gap would be systematically larger than ours; this is
probably related to the neglect of spin-orbit coupling in their calculation.
It can be seen that the size dependence of our uncorrected $E_g$
shows a good qualitative agreement with experiment.
The calculated values are somewhat smaller, but they are simply based on
the parameters of Ref. \onlinecite{kobayashi} for bulk CdTe, which we prefer
not to modify in any arbitrary manner.
There is also some uncertainty related to the parameters concerning
the H atoms.
As for the Coulomb correction, it is rather difficult to evaluate
consistently within the TB model. But we expect it to be significantly smaller
than that estimated from the EMA used in the calculations cited above,
because the TB carrier wave functions are spatially more extended
(see Fig.~5).

Figure 5 shows the shell-wise radial distribution of the carrier
probability
density for the LUS (electron) and HOS (hole) states in a QD of diameter
58.67 {\AA}.
The shell probability density for an eigenstate $\vert{\Psi}\rangle$ is directly
obtained from the
diagonalization of the TB Hamiltonian, by summing the local density
$P({\bf R}_b)\equiv\sum_{i,\sigma}|<{\bf R}_{b},i,\sigma|\Psi>|^2$
over the sites ${\bf R}_{b}$ on the given shell.
We see that, in contrast with the smooth EMA envelope functions, the
TB radial distributions are oscillatory. It is interesting to note that
the LUS (HOS) is indeed preponderent at the cation (anion) shells.
In the present case the cation (anion) shells account for 79\% (76\%)
of the LUS (HOS).
Another reason for the oscillatory behavior is the discontinuous
variation of the number of atoms on a shell. All the oscillations are, of
course, smoothed over in the EMA. It is not easy to make a direct comparison.
However, it can be verified that our TB charge distributions have a greater
radial extension than the EMA envelope functions.

Finally, in order to reexamine the electric-dipole selection rules for the
fundamental
interband transition,
we present an analysis of the orbital symmetry of the full wave functions
in the TB model.
We proceed as follows. By setting the spin-orbit interaction $H_{\text{SO}}=0$,
we first diagonalize $H_0$ in the spin-degenerate atomic orbital basis.
The resulting eigenvalues are then classified according to the single-valued
representations $\Gamma_i$ ($i=1 - 5$). As the second step, we diagonalize
the same Hamiltonian in the spin-orbit coupled basis: the same eigenvalues
are now classified in terms of the double-valued representations
$\Gamma_i$ ($i=6 - 8$), which arise from
the direct products of $\Gamma_6 $ with $\Gamma_i$ ($i=1 - 5$).
By consulting the multiplication table of the group $T_d$,
we see that there are three different types of $\Gamma _8$
corresponding to the three orbital symmetries $\Gamma _3$, $\Gamma _4$ and
$\Gamma _5$. Thus, the eigenstates $\Gamma
_8^{0}$ of $H_0$ can be identified as $\Gamma _8^{0}(\Gamma_3)$,
$\Gamma _8^{0}(\Gamma_4)$ or $\Gamma _8^{0}(\Gamma_5)$.
Finally, a given eigenstate
$\Gamma _8$ of the full Hamiltonian, for example the HOS,
can be written as a linear combination of all the
`unperturbed' states
$\Gamma _8^{0}(\Gamma_i)$ $(i=3-5)$.
By calculating the projections and summing the probabilities over
all states arising from a given orbital symmetry $\Gamma _i$, we can deduce
the fractional contribution of that symmetry in
the HOS. Similarly, an eigenstate $\Gamma _6$ such as the LUS can be analyzed
in terms of the orbital symmetries $\Gamma _1$ and $\Gamma _4$.

The numerical results for the CdTe QD's are as follows. The LUS is found to be
almost pure $\Gamma _1$ for all sizes: 99.9\% for $D=12.35$ {\AA}
and 99.7\% for $D=58.67$ {\AA}. The results for the HOS ($\Gamma _8$)
are more interesting; they are shown in Table~\ref{tab2}.
Note that our numerical calculation was limited to
the contributions from the nine topmost `unperturbed' valence
states
for each of the single-valued representations, because of the huge size of
the data files involved.
This
is, however, adequate for identifying the majority component in all
cases. We find that the $\Gamma _3$ contribution is always small. On the
other hand, as the QD size increases the {\it dominant} orbital symmetry of
the HOS changes from $\Gamma_ 4$ to $\Gamma_ 5$. Thus, for $D\geq 17.8$ {\AA}
the fundamental interband transition is certainly dipole-allowed.
In Fig.~2 we notice that the second highest valence state is also of $\Gamma_8$
symmetry and lies very close to the HOS. We, therefore, analyzed its orbital
symmetry as well; the results are presented inside parentheses
in Table~\ref{tab2}. It is interesting to observe that its dominant orbital
symmetry also changes with increasing size, but from $\Gamma_ 5$ to $\Gamma_
4$, in a manner opposite to the HOS.
Anyway, the QD size where this second level approaches the HOS is large
enough so that no further `symmetry cross-over' is expected, because
the HOS in the bulk is indeed $\Gamma_5$-like. To sum up,
the fundamental transition is {\it almost}
dipole-forbidden for $D\leq 13.2$ {\AA}, but certainly dipole-allowed
for $D\geq 17.8$ {\AA}.
Clearly, our results contradict the prediction of
Lefevre {\it et al.}\cite{lefevre}.
They found that
in the three-band EMA the HOS (hole ground state) in a CdTe QD, regardless of
the size, has a $p$-type envelope function, making it dipole-forbidden for
optical transitions with the $s$-type LUS.
Let us point out that although our analysis is inadequate for calculating the
actual
transition probabilities, it does allow an unambiguous identification of
the symmetry forbidden transitions.

\section{concluding remarks}

 We have presented a TB calculation of the electronic structure of a
roughly spherical semiconductor QD of tetrahedral symmetry. It is based on
the $sp^3s^*$ empirical TB model\cite{vogl}
that accounts for the band structure of bulk semiconductors.
The use of group theory allows not only a substantial reduction in size of
the Hamiltonian matrix to diagonalize, but also a symmetry analysis
of the
eigenstates. Our treatment represents a generalization of the previous work
by Ren and Dow\cite{ren} to binary compounds and a finite
spin-orbit interaction. Here we have applied the model to the case of CdTe.
Through a careful analysis of the DOS spectrum
 we have explicitly passivated
the surface dangling-bond states.
The calculated one-electron bandgap as a function of size shows a reasonable
agreement with the available experimental values of the optical bandgap.
We find that,
regardless of the QD size, the valence band HOS and the conduction band LUS
belong to
$\Gamma_8$ and $\Gamma_6$ symmetries, respectively, as in the bulk
semiconductor.
However, an analysis of the orbital symmetry reveals that, while the LUS
is almost pure $\Gamma_1$-like for all sizes, the {\it dominant} orbital
symmetry of the HOS changes
from $\Gamma _4$ for $D\leq 13.2$ {\AA} to $\Gamma_5$
for $D\geq 17.8$ {\AA}. Thus, except for very small QD's,
the electric dipole transitions remain
allowed between the HOS and the LUS, in contradiction with the three-band EMA
result reported previously.

\acknowledgments

This work was supported in part by the spanish DGICyT under
 contract no. PB95-0797.

\begin{table}
\begin{tabular}{|c|c|c|c|c|c|}
$D$ (\AA)&$N$&$N_{\text{H}}$&Last shell
 & $E$(HOS) & $E$(LUS)\\
    \hline\hline
$10.34$ & $17$	 & $36$  & cation & $-1.101$ & $2.909$	\\ \hline
$12.35$ & $29$	 & $36$  & anion  & $-1.083$ & $2.113$	 \\ \hline
$13.15$ & $35$	 & $36$  & cation & $-0.969$ & $2.427$	\\ \hline
$17.81$ & $87$	 & $76$  & anion  & $-0.719$ & $1.951$	 \\ \hline
$24.95$ & $239$  & $196$ & cation & $-0.393$ & $2.059$	\\ \hline
$32.43$ & $525$  & $276$ & anion  & $-0.303$ & $1.721$ \\ \hline
$43.08$ & $1231$ & $460$ & cation & $-0.177$ & $1.775$	\\ \hline
$58.67$ & $3109$ & $852$ & cation & $-0.104$ & $1.696$
\end{tabular}
\caption{Some important features of the QD's analyzed.
 The first column gives the diameter. The second and third
 columns give the numbers of semiconductor and hydrogen atoms, respectively.
The fourth column specifies the atom type of the
terminating shell.
The last two columns show
 the calculated HOS and LUS energies in eV.}
\label{tab1}
\end{table}

\begin{table}
\begin{tabular}{|c|c|c|c|}
$D$ (\AA)&
 $\Gamma_{3}$ & $\Gamma_{4}$ & $\Gamma_{5}$ \\	     \hline\hline
$12.35$ & $3.88$($6.85$) & $84.15$($17.07$) & $11.57$($75.92$)	 \\ \hline
$13.15$ & $4.87$($4.48$) & $90.26$($33.84$) & $3.78$($49.03$)	 \\ \hline
$17.81$ & $4.25$($0.54$) & $22.04$($80.10$) & $68.74$($18.16$)	 \\ \hline
$24.95$ & $0.10$($0.20$) & $0.11$($87.84$)  & $96.54$($1.26$)	 \\ \hline
$32.43$ & $1.43$($0.60$) & $33.27$($63.65$) & $54.57$($31.30$)	 \\ \hline
$43.08$ & $1.81$($0.99$) & $26.35$($67.71$) & $58.81$($26.43$)	 \\ \hline
$58.67$ & $1.68$($0.77$) & $26.91$($65.44$) & $56.70$($27.58$)
\end{tabular}
\caption{Size dependence of the percentage contributions of different orbital
symmetries in the
HOS. Inside parentheses we show the corresponding values for the nearest state
  in energy (also $\Gamma_8$ ),
  which is almost degenerate for
  $D>30$ {\AA\@}.}
\label{tab2}
\end{table}

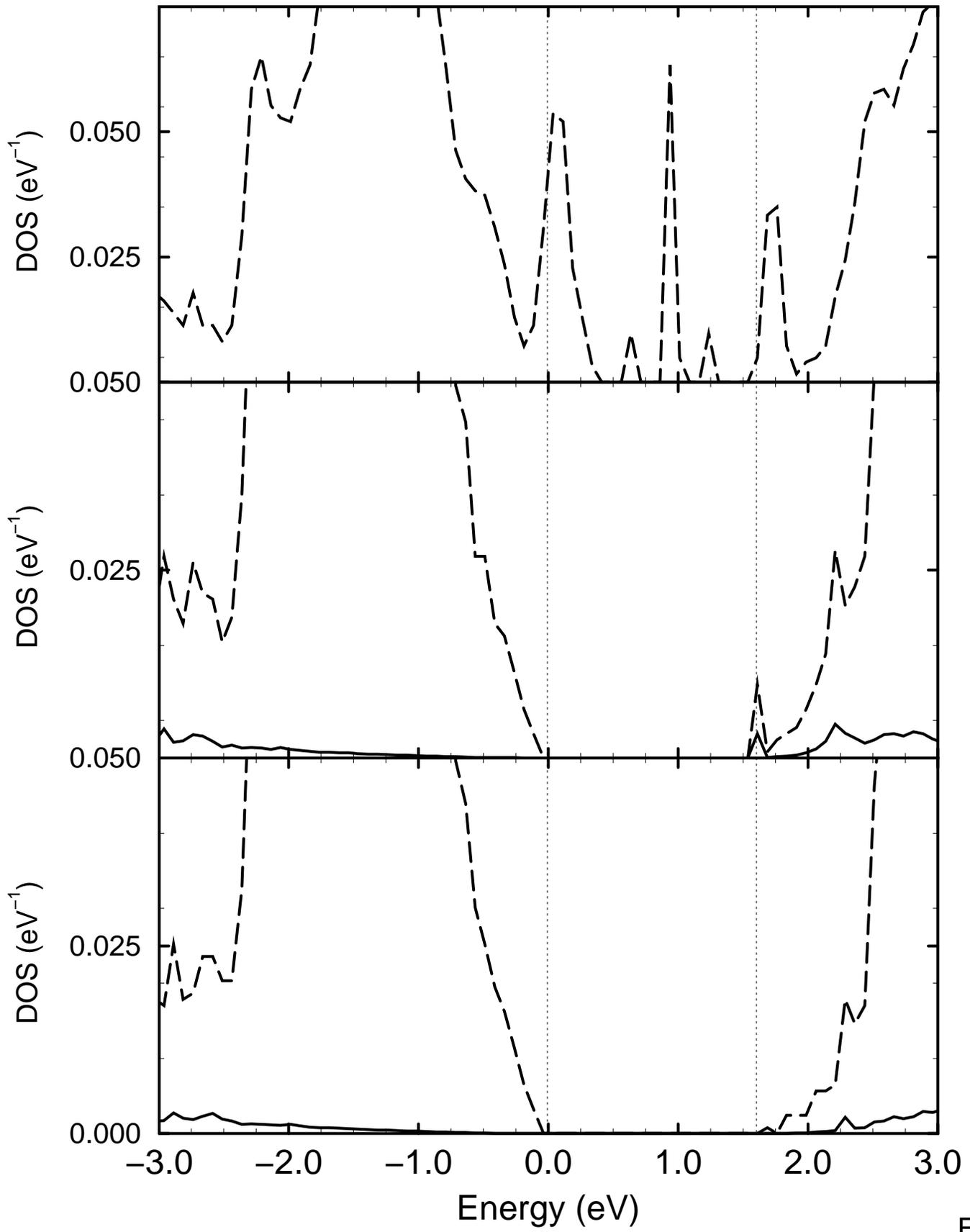
\begin{figure}
\caption{
Total density of states (dashed curve) and its hydrogenic part (solid curve)
 for $D=58.67$ \AA\@: (a) No passivation, (b) H passivation with the
 bond lengths $d_{\text{Te-H}}=1.67$ and $d_{\text{Cd-H}}=1.71$ \AA\@, (c) H
passivation with the bond lengths
$d_{\text{Te-H}}=1.54$ and $d_{\text{Cd-H}}=1.58$ \AA\@ }
\label{fig1}
\end{figure}

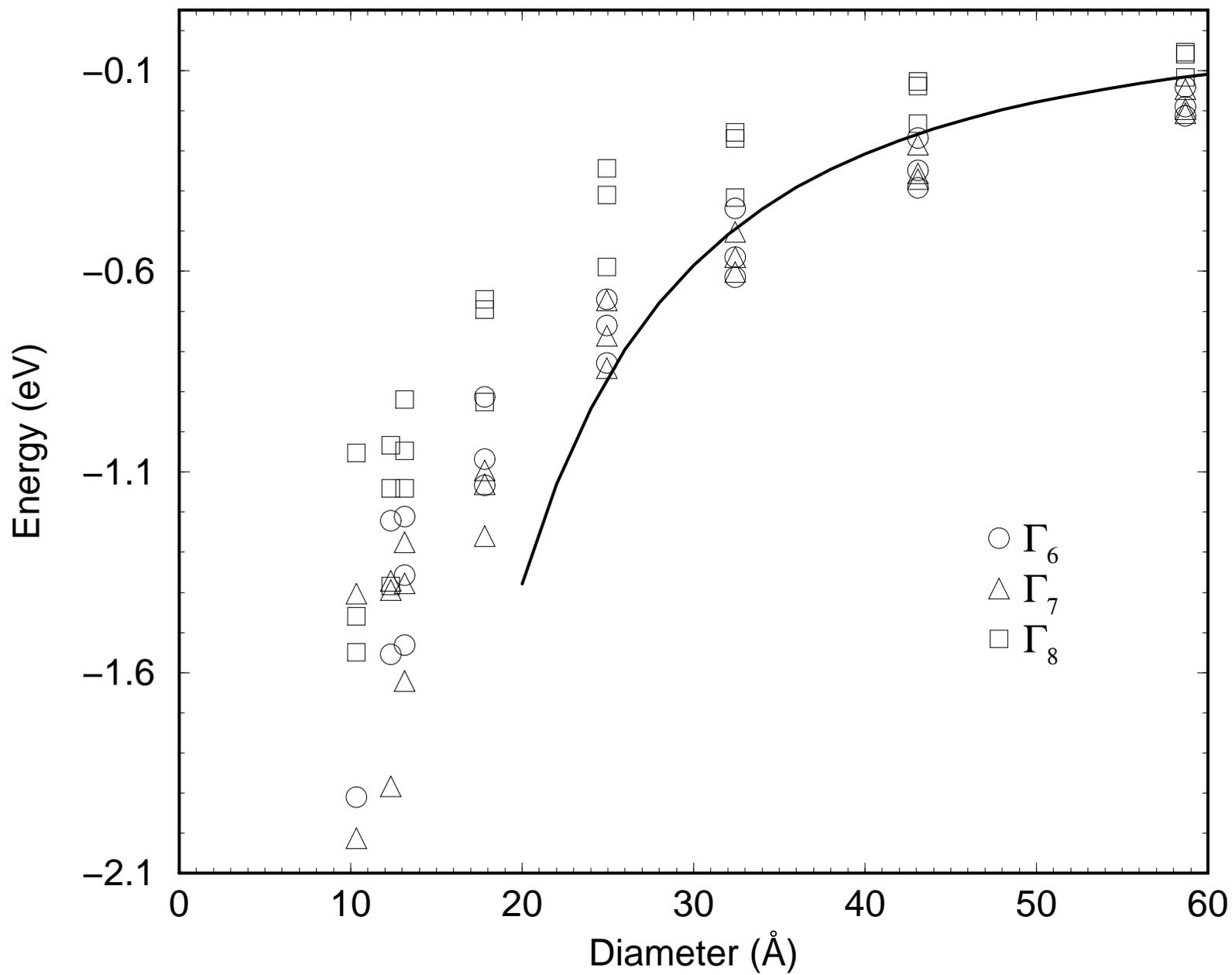
\begin{figure}
\caption{Energies and symmetries of the ten highest occupied (valence band)
levels
{\it versus} the QD diameter.
The solid curve shows the three-band EMA result from Ref.
 \protect\onlinecite{lefevre}
for the highest level.}
\label{fig2}
\end{figure}

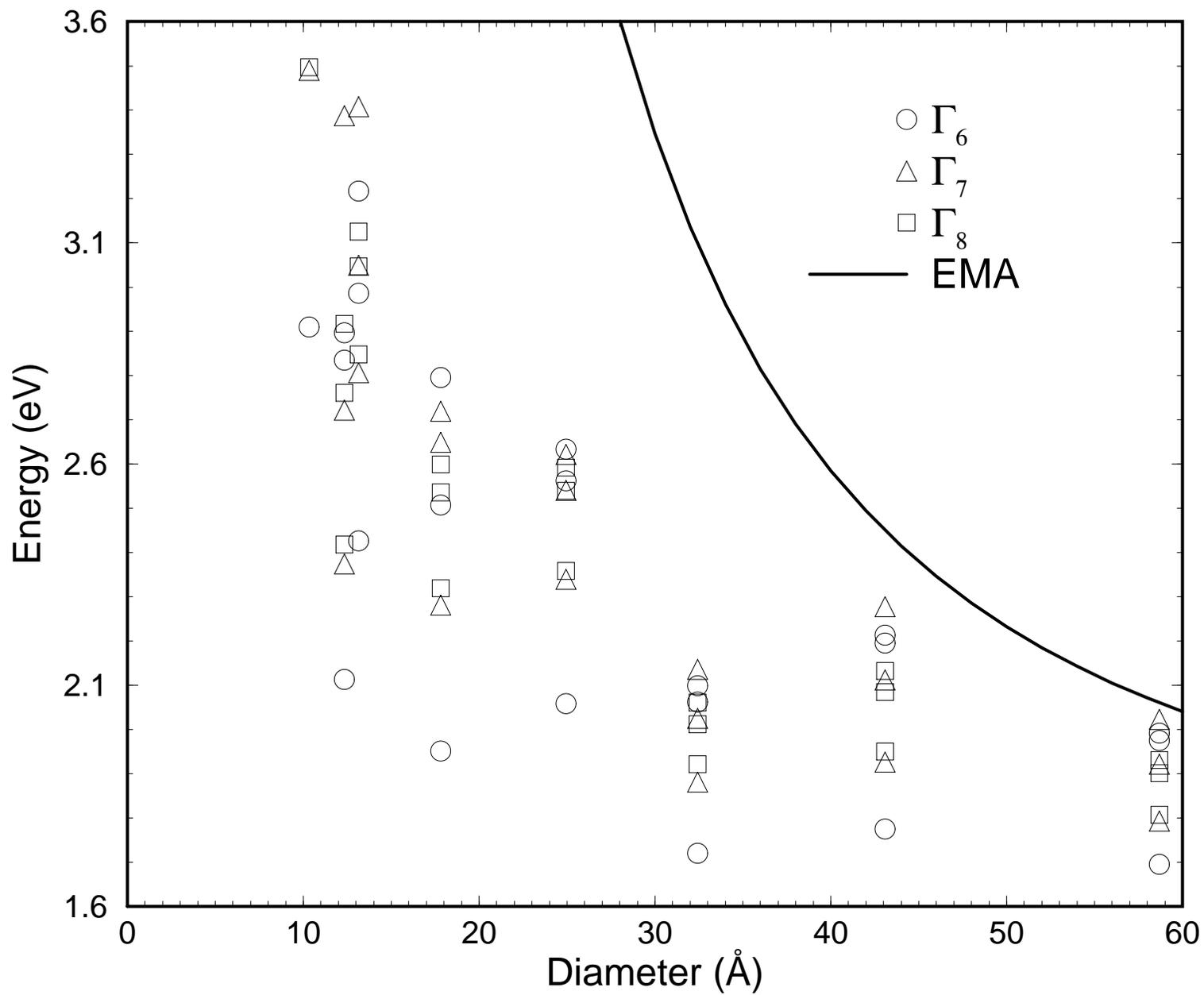
\begin{figure}
\caption{Energies and symmetries of the ten lowest unoccupied (conduction)
levels {\it versus} the QD diameter.
The solid curve is the EMA result for the lowest level.}
\label{fig3}
\end{figure}

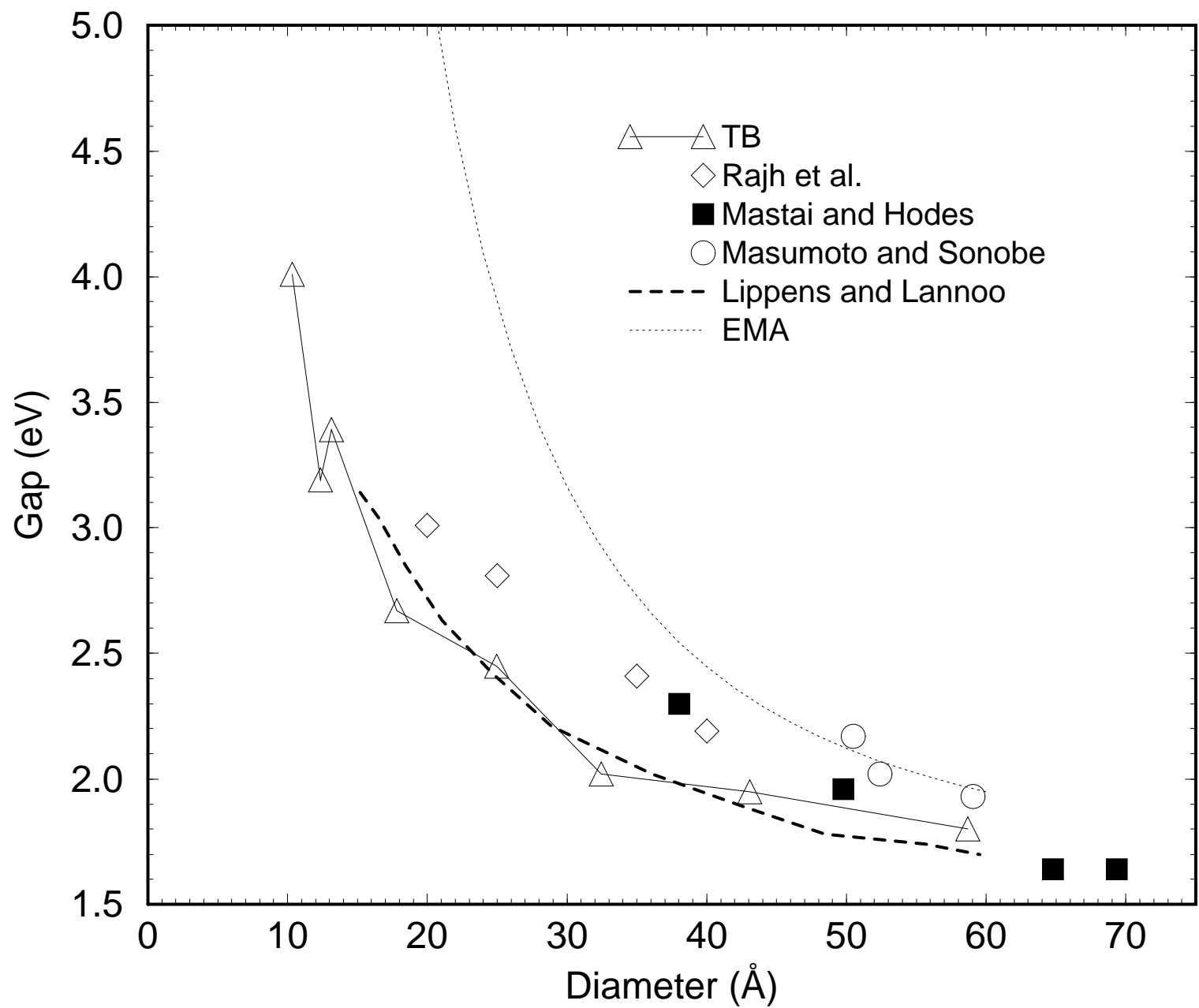
\begin{figure}
\caption{Size dependence of the energy gap: Results of the present
calculation are shown as open triangles connected by solid lines.
Experimental data for the optical gap are from Refs. \protect\onlinecite{rajh}
(open diamonds), \protect\onlinecite{mastai}
 (filled squares), and \protect\onlinecite{masumoto} (open circles).
Previous calculations in the TB model (Ref.
\protect\onlinecite{lippens}) and the EMA
(Ref. \protect\onlinecite{lefevre})
are shown as the dashed and dotted curves, respectively.}
\label{fig4}
\end{figure}

\begin{figure}
\caption{Shell-wise radial distribution of the carrier probability density
in the HOS (a) and LUS (b) for $D=58.67$ {\AA}.}
\label{fig5}
\end{figure}

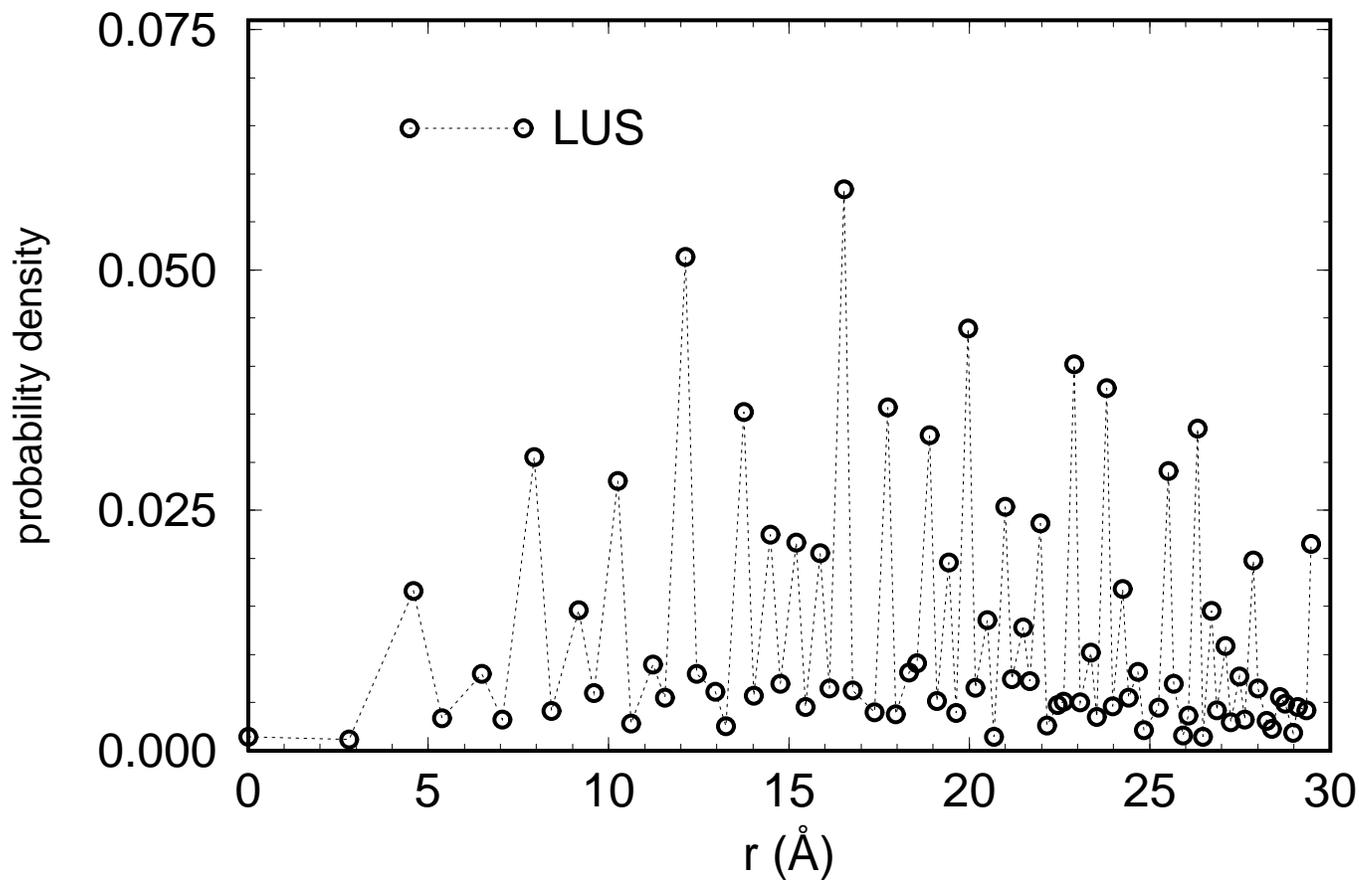
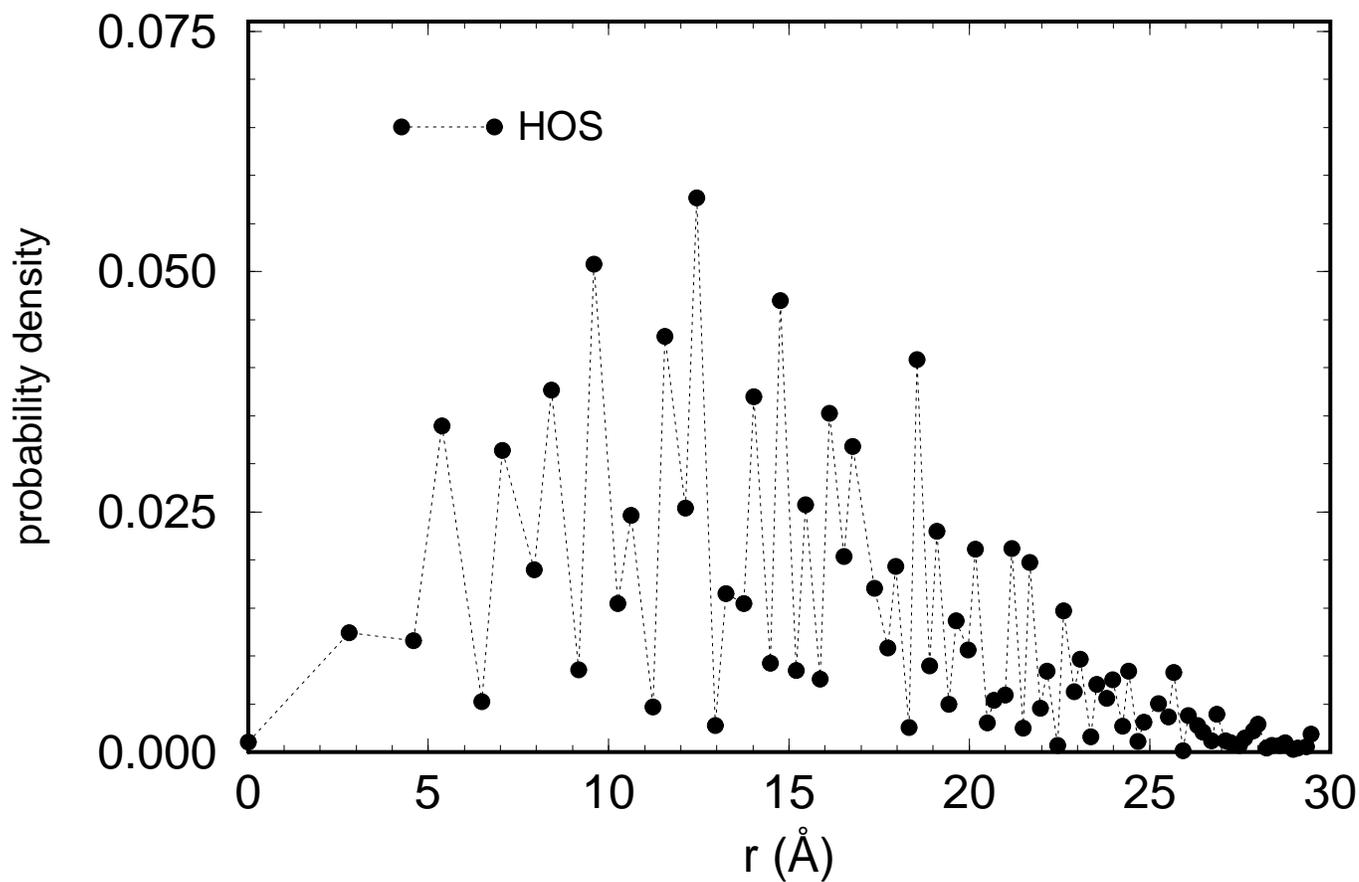
\end{document}